\documentclass[12pt]{iopart}
\usepackage{amssymb}
\usepackage{graphicx}
\usepackage{color}
\eqnobysec

\def\keywords#1{\vspace{10pt}
     \begin{indented}
     \item[]\rm Keywords: #1\par
     \end{indented}}

\def\be{\begin{equation}}
\def\ee{\end{equation}}
\def\bea{\begin{eqnarray}}
\def\eea{\end{eqnarray}}
\def\BD{\mathbf{\Delta}}
\def\BE{\mathbf{E}}

\def\BI{\mathbf{I}}
\def\CP{{\cal P}}
\def\CQ{{\cal Q}}
\def\CR{{\cal R}}
\def\CS{{\cal S}}

\begin{document}
\jl{1}

\title[Random walk on the fully-connected lattice]{Probability distribution of the number of distinct sites visited by a random walk on the finite-size fully-connected lattice}

\author{Lo\"\i c Turban}

\address{Groupe de Physique Statistique, D\'epartement P2M, Institut Jean Lamour,\\ Universit\'e de Lorraine, CNRS (UMR 7198), 
Vand\oe uvre l\`es Nancy Cedex, F-54506, France} 

\ead{loic.turban@univ-lorraine.fr}

\begin{abstract}
The probability distribution of the number $s$ of distinct sites visited up to time $t$ by a random walk on the fully-connected lattice with $N$ sites is first obtained by solving the eigenvalue problem associated with the discrete master equation. Then, using generating function techniques, we compute the joint probability distribution of $s$ and $r$, where $r$ is the number of sites visited only once up to time $t$. Mean values, variances and covariance are deduced from the generating functions and their 
finite-size-scaling behaviour is studied. Introducing properly centered and scaled variables $u$ and $v$ for $r$ and $s$ and working in the scaling limit ($t\to\infty$, $N\to\infty$ with $w=t/N$ fixed) the joint probability density of $u$ and $v$ is shown to be a bivariate Gaussian density. It follows that the fluctuations of $r$ and $s$ around their mean values in a finite-size system are Gaussian in the scaling limit. The same type of finite-size scaling is expected to hold on periodic lattices above the critical dimension $d_{\rm c}=2$.
\end{abstract}

\keywords{random walk, fully-connected lattice, visited sites}
\pacs{02.50.-r, 05.40.-a}


\section{Introduction} 
The study of random walks and Brownian motion, associated with the names of Bachelier~\cite{bachelier00}, Einstein~\cite{einstein05}, Smoluchowski~\cite{smoluchowski06} and Langevin~\cite{langevin08} at the beginning of the last century, still remains an active field of research with many applications in various domains of physics~[5--9], mathematics~[10--12], physical chemistry~\cite{vankampen81,rice85}, biology~[15--17], economics~\cite{bouchaud05}, etc. 

The discrete random walk on a periodic lattice was introduced  by Polya~\cite{polya21} who showed that, with probability 1, a random walker returns infinitely often to the origin in dimension $d\leq2$ and escapes to infinity when $d\geq3$. The walk is said to be recurrent in the first case and transient in the second. A quantity of interest for applications is the mean number 
of distinct sites, $\overline{s_\infty}(t)$, visited by a random walker on an infinite periodic lattice up to time $t$. In $1d$ (one dimension) $\overline{s_\infty}(t)$ evidently grows as $\sqrt{t}$. The asymptotic behaviour in higher dimensions was first obtained by Dvoretzky and Erd\"os~\cite{dvoretzky51} with
\bea
&\overline{s_\infty}(t)= \pi t/\ln (t)+O\left[\frac{t\ln(\ln t)}{\ln^2 t}\right]\,,\qquad &d=2\quad {\rm (square\ lattice)}\,,\nonumber\\
&\overline{s_\infty}(t)=\gamma_3 t+O(\sqrt{t})\,,\qquad &d=3\,,\nonumber\\
&\overline{s_\infty}(t)=\gamma_4 t+O(\ln t)\,,\qquad &d=4\,,\nonumber\\
&\overline{s_\infty}(t)=\gamma_d t+O(1)\,,\qquad &d>4\,.
\label{sinft}
\eea
Thus $d=2$ plays the role of a critical dimension $d_{\rm c}$ above which the time exponent of the leading contribution stays constant~\footnote[1]{This is true only when $d>4$ for the sub-leading terms. Note that the fractal dimension of the intersection of two random walks vanishes when $d\geq4$~\cite{mandelbrot82}}. 
Exact values of the amplitudes $\gamma_d$ and sub-dominant contributions were later calculated for different lattices in $3d$~\cite{vineyard63,montroll65}. 
The mean number of distinct sites visited exactly $k$ times was also examined for $d\leq3$~\cite{montroll65}. For 
$k=1$ the following results were obtained~\cite{montroll65,erdos60}:
\bea
&\overline{r_\infty}(t>1)=2\,, \qquad &d=1\,,\nonumber\\
&\overline{r_\infty}(t)\sim t/(\ln t)^2\,,\qquad     &d=2\,,\nonumber\\
&\overline{r_\infty}(t)\sim t\,,\qquad     &d=3\,.
\label{rinft}
\eea
More recently, the mean values of different local observables associated with the geometry of the set of visited sites have been studied in $2d$~\cite{vanwijland97a} and $d>2$~\cite{vanwijland97b}.

Instead of considering a single random walk, one may generalize to the case of $\nu$ statistically independent random walks. Such studies were first concerned with the properties of first passage times~[27--30].
The study of the mean number of distinct sites visited by $\nu$ random walkers, $\overline{s_{\infty,\nu}}(t)$, was initiated in~\cite{larralde92} where asymptotic expressions for $\nu$ large were obtained. In~\cite{yuste99,yuste00a} some corrections to~\cite{larralde92} were given and sub-dominant contributions were evaluated. 

With several random walkers, another quantity of interest is the number of common sites visited up to time $t$. This study was initiated in~\cite{majumdar12} where the mean value and the variation with $d$ of the long-time behaviour was examined. It was later generalized by considering two walkers starting from different points~\cite{tamm14}.

The number of distinct sites visited up to time $t$ can be written as $s_\infty(t)=\sum_{k=1}^t\sigma_k$ where $\sigma_k$ is an indicator such that $\sigma_k=1$ when a new site is visited at time $k$ and $\sigma_k=0$ otherwise. Thus $s_\infty(t)$ is a sum of random variables which are neither independent, nor identically distributed. Nevertheless it has been shown that for transient walks ($d\geq3$) the deviations from the mean display Gaussian fluctuations~\cite{jain68,jain71}. 

More recent results concern recurrent random walks. In $1d$ non-trivial exact probability distributions for the number of distinct and common sites visited by $\nu$ independent walkers have been obtained~\cite{kundu13}. In $2d$ and $3d$ it has been shown that, at long time, the deviations from the mean of different local observables associated with the set of visited sites are proportional to a single universal random process which is non--Gaussian in $2d$ and Gaussian in $3d$~\cite{vanwijland97a,vanwijland97b}. 

\begin{figure}[!t]
\begin{center}
\includegraphics[width=10cm,angle=0]{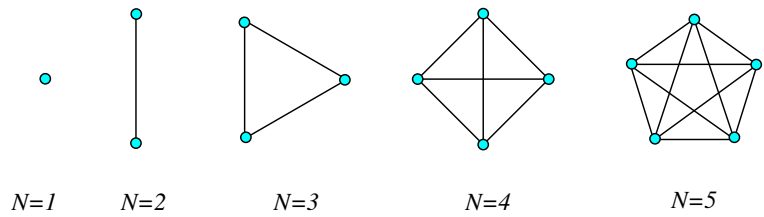}
\end{center}
\vglue -.0cm
\caption{\label{fig-1} The fully-connected lattice (complete graph) with $N$ sites can be embedded in a Euclidean space of dimension $d=N-1$.
}
\end{figure}

This paper is concerned with the statistics of the number of sites visited by a random walker up to time $t$ on a finite-size fully-connected lattice (see figure~\ref{fig-1}). 
The fully-connected lattice with $N$ sites can be embedded in a Euclidean space of $N-1$ dimensions. It can  be also considered as a finite-size system embedded in an infinite-dimensional space where the thermodynamic limit $N\to\infty$ can be taken. Such lattices have been used in the theory of phase transitions~[39--41] and in models of the nucleus~\cite{lipkin65} to obtain exact solutions.  In the thermodynamic limit the critical behaviour is that of a system above its upper critical dimension $d_{\rm c}$, i.e., mean-field like. Since the system lives in a space with $d=\infty$ and is characterized by a number of sites $N$, instead of a length $L$, the formulation of finite-size scaling is not standard~[43--45].

Our main results can be summarized as follows. We have obtained exact expressions for the probability distribution of the number $s$ of distinct sites visited by the random walk up to time $t$
\be
S_N(s,t)=\frac{N^{\underline{s}}}{N^t}\,{t\brace s}\,,
\ee
the probability distribution of the number $r$ of sites visited once up to time $t$
\be
R_N(r,t)=\frac{1}{N^t}\sum_{k=r}^t(-1)^{k+r}{t\choose k}{k\choose r}(N-k)^{t-k}N^{\underline{k}}\,,
\ee
and their joint probability distribution
\be
P_N(r,s,t)=
\frac{N^{\underline{s}}}{N^t}\sum_{k=r}^s(-1)^{k+r}{t\choose k}{k\choose r}{t-k\brace s-k}\,.
\ee
In these expressions $n^{\underline{m}}$ is a falling factorial power~(\cite{graham94} p 47)
and ${n\brace m}$ is a Stirling number of the second kind~\cite{stirling49}.

In the scaling limit ($t\to\infty$, $N\to\infty$ with $w=t/N$ fixed) $R_N(r,t)$ and $S_N(s,t)$ lead to centered Gaussian probability densities in the variables 
\be
u=\frac{r-\overline{r_N(t)}}{N^{1/2}}\,,\qquad v=\frac{s-\overline{s_N(t)}}{N^{1/2}}\,,
\ee 
with the following mean values for $r$ and $s$ at time $t$
\be
\overline{r_N(t)}=t\,\rme^{-w}\,,\qquad  \overline{s_N(t)}=t\,\frac{1-\rme^{-w}}{w}\,,
\ee
while $P_N(r,s,t)$ leads to a bivariate Gaussian density. The elements of the covariance matrix are given by:
\be\fl
\overline{\Delta u^2}\!=\!w\!\left[\rme^{-w}\!-\!(w^2\!-\!w\!+\!1\!)\rme^{-2w}\right]\!,\quad
\overline{\Delta v^2}\!=\!\rme^{-w}\!-\!(w\!+\!1)\rme^{-2w}\!,\quad  
\overline{\Delta u\Delta v}\!=\!w^2\rme^{-2w}\!.
\ee
A similar Gaussian finite-size behaviour is expected on periodic lattices above the critical dimension $d_{\rm c}=2$.

The outline of the paper is as follows. In section 2, we solve the eigenvalue problem associated with the discrete master equation governing the probability distribution $S_N(s,t)$ of the number of distinct sites $s$ visited by a random walk up to time $t$. The probability distribution is obtained and generalized to the case of $\nu$ independent walkers. Section 3 is devoted to the study of the probability distribution $R_N(r,t)$ of the number of sites $r$ visited once up to time $t$. Actually, we first solve the master equation for the joint probability distribution $P_N(r,s,t)$, from which $R_N(r,t)$ is deduced, using a generating function technique. Next, in section 4, mean values, variances and covariance of $r$ and $s$ are computed and their finite-size scaling behaviour is examined.  Finally, in section 5, using properly scaled variables, the probability densities are obtained in the scaling (continuum) limit. Details of the calculations are given in five appendices.

\section{Total number of distinct sites visited by the random walk}
\subsection{Discrete master equation}
The random walk we consider takes place on the fully-connected lattice (complete graph) with $N$ sites. At each time step,
with probability $1/N$, the walker either remains on the same site or jumps to any of the $N-1$ other sites. We shall study the probability distribution $S_N(s,t)$ of the number $s$ of distinct sites visited by the walker up to time~$t$. It satisfies the following recurrence relation
\be
S_N(s,t)=\frac{s}{N}\,S_N(s,t-1)+\frac{N-s+1}{N}\,S_N(s-1,t-1)\,,
\label{rec-1}
\ee
with the boundary condition $S_N(s<0,t)=0$. The first (second) term on the right corresponds to a step on one of the $s$ (respectively $N-s+1$) sites already (not yet) visited at time $t-1$. The walk starts at some arbitrary origin at time $t=1$. Thus the initial condition can be written as $S_N(s,1)=\delta_{s,1}$ or alternatively, according to \eref{rec-1},  $S_N(s,0)=\delta_{s,0}$. The stationary solution is given by $S_N(s,t)=\delta_{s,N}$.

\subsection{Eigenvalue problem}
Let us introduce the column state vector $|S_N(t)\rangle$ with components $S_N(s,t)$, $s=1,\ldots,N$. The master equation~\eref{rec-1} can be rewritten in matrix form as
$|S_N(t)\rangle=\mathsf{T}\,|S_N(t-1)\rangle$ where $\mathsf{T}$ is the transition matrix of the Markov chain given by:
\be
\mathsf{T}=\left(\begin{array}{cccccc}
\frac{1}{N}   &0	  &0	            &0           &0           &0          \\
\frac{N-1}{N} &\frac{2}{N}&0	            &0           &0           &0          \\
	      &  \ddots   &\ddots           &            &            &           \\
0	      &0          &\frac{N-s+1}{N}  &\frac{s}{N} &0           &0          \\
              &           &                 &\ddots      &\ddots      &           \\ 
0	      &0	  &0	            & 0	         &\frac{1}{N} &1          \\
\end{array}\right)\,.
\label{tmatrix}
\ee
The eigenvalue problem $(\mathsf{T}|v^{(k)}\rangle=\lambda_k|v^{(k)}\rangle$ leads to the following system of equations
\be
\frac{N-s+1}{N}v_{s-1}^{(k)}+\left(\frac{s}{N}-\lambda_k\right)v_s^{(k)}=0\,,\qquad s=1,\ldots,N\,,
\label{eigen-1}
\ee
with $v_0^{(k)}=0$. It is easy to verify that the solution is given by
\be\fl
\lambda_k=\frac{k}{N}\,,\quad v_s^{(k)}=\left\{
\begin{array}{lll}
0& {\rm when} &s<k\\
(-1)^{s-k}{N-k\choose s-k}v_k^{(k)}&{\rm when}& s\geq k\\
\ms
\end{array}\right.\,,\quad k=1,\ldots,N\,,
\label{eigen-2}
\ee 
where the $v_k^{(k)}$ are left undetermined and depend on the initial state. 

\subsection{Probability distribution}
Let us consider a walk starting from some arbitrary site at $t=1$ so that $S_N(s,1)=\delta_{s,1}$. With the following choice for the initial state 
\be
|S_N(1)\rangle=\sum_{k=1}^N|v^{(k)}\rangle\,, \qquad v_k^{(k)}={N-1\choose k-1}\,,
\label{sn1}
\ee
one obtains 
\bea\fl
S_N(s,1)&=\sum_{k=1}^sv_s^{(k)}=\sum_{k=1}^s(-1)^{s-k}{N-k\choose s-k}{N-1\choose k-1}
={N-1\choose s-1}\sum_{k=1}^s(-1)^{s-k}{s-1\choose k-1}\nonumber\\
\fl&={N-1\choose s-1}\sum_{l=0}^{s-1}(-1)^l{s-1\choose l}=\delta_{s,1}\,,
\label{sns-1}
\eea
as required for this initial state.

The probability distribution at later times is given by the state vector
\be
|S_N(t)\rangle=\mathsf{T}^{t-1}|S_N(1)\rangle=\sum_{k=1}^N\lambda_k^{t-1}|v^{(k)}\rangle\,,
\label{sn}
\ee
so that
\bea
\fl S_N(s,t)&={N-1\choose s-1}\sum_{k=1}^s(-1)^{s-k}{s-1\choose k-1}\left(\frac{k}{N}\right)^{t-1}\!\!\!\!\!\!
=\frac{N^{\underline{s}}}{N^t(s-1)!}\sum_{k=1}^s(-1)^{s-k}{s-1\choose k-1}k^{t-1}\nonumber\\
\fl&=\frac{N^{\underline{s}}}{N^t}\,\frac{1}{s!}\sum_{k=0}^s(-1)^{s-k}{s\choose k}k^t=\frac{N^{\underline{s}}}{N^t}\,{t\brace s}\,.
\label{snst-1}
\eea
Here $N^{\underline{s}}=N(N-1)\cdots (N-s+1)$ and 
\be
{t\brace s}=\frac{1}{s!}\sum_{k=0}^s(-1)^{s-k}{s\choose k}k^t
\label{stir-1}
\ee
is a Stirling number of the second kind~(\cite{graham94}~p~257; see table 1). Note that the term $k=0$ does not contribute in \eref{snst-1} where $t>0$. 

When the forward-difference operator $\BD$, such that $\BD f(\eta)=f(\eta+1)-f(\eta)$, is applied $s$ times to $f(\eta)$ one obtains~(\cite{graham94} p 188)
\be
\BD^s f(\eta)=\sum_{k=0}^s(-1)^{s-k}{s\choose k}f(\eta+k)\,,
\label{fdif}
\ee 
which follows from the relation $\BD=\BE-\BI$ where $\BE$ is the shift operator, 
such that $\BE f(\eta)=f(\eta+1)$, and $\BI$ the identity operator.
Using \eref{fdif}, the Stirling number of the second kind in equation~\eref{stir-1} can be rewritten as
\be
{t\brace s}=\frac{1}{s!}\left.\BD^s \eta^t\right|_{\eta=0}
\label{stir-2}
\ee
and the probability distribution in~\eref{snst-1} takes the following form:
\be
S_N(s,t)=\frac{N^{\underline{s}}}{N^ts!}\left.\BD^s \eta^t\right|_{\eta=0}
=\frac{1}{N^t}{N\choose s}\left.\BD^s \eta^t\right|_{\eta=0}\,.
\label{snst-2}
\ee
This last expression is used in appendix~A to calculate the generating function
\be
\CS_N(y,z)=\left[1+y\left(\rme^{z/N}-1\right)\right]^N\,.
\label{snyz}
\ee
This generating function is ordinary in $y$ but exponential in $z$.

The behaviour of $S_N(s,t)$ when both $t$ and $N$ are large, will be analysed in section~5.

\begin{table}
\caption{\label{table:1} Stirling number of the second kind: the table is constructed using the recursion which follows from equations~\eref{rec-1} and~\eref{snst-1},
${t\brace s}=s{t-1\brace s}+{t-1\brace s-1}$ with ${t\brace 0}=\delta_{t,0}$ and ${t\brace s}=0$ when $s>t$.}
\begin{indented}
\item[]\begin{tabular}{@{}lrrrrrrrrrr}
\br\ms
t\verb"\"s & 0  & 1  & 2  & 3  & 4  & 5  & 6  & 7 \\
\mr
0          & 1  & 0  &    &    &    &    &    &   \\ 
1          & 0  & 1  & 0  &    &    &    &    &   \\
2          & 0  & 1  & 1  & 0  &    &    &    &   \\
3          & 0  & 1  & 3  & 1  & 0  &    &    &   \\
4          & 0  & 1  & 7  & 6  & 1  & 0  &    &   \\
5          & 0  & 1  & 15 & 25 &10  & 1  & 0  &   \\
6          & 0  & 1  & 31 & 90 &65  & 15 & 1  & 0 \\
\br
\end{tabular}
\end{indented}
\end{table}

\subsection{The case of $\nu$ independent walkers}

The evolution of $s$ with $\nu$ independent walkers is the same in one time step as for a single walker in 
$\nu$ time steps. When the $\nu$ walkers start from arbitrary sites at $t=1$ (initial condition $S_{N,\nu}(s,0)=\delta_{s,0}$) the probability distribution is simply obtained by changing $t$ into $\nu t$ in equation~\eref{snst-1} so that
\be
S_{N,\nu}(s,t)=\frac{N^{\underline{s}}}{N^{\nu t}}\,{\nu t\brace s}\,.
\label{snnut-1}
\ee
If the $\nu$ walkers start from the same origin at $t=1$ (initial condition $S_{N,\nu}(s,1)=\delta_{s,1}$) $\mathsf{T}$ must be replaced by $\mathsf{T}^\nu$ in equation~\eref{sn} which amounts to change $t$ 
into $1+\nu(t-1)$ leading to
\be
S_{N,\nu}(s,t)=\frac{N^{\underline{s}}}{N^{1+\nu (t-1)}}\,{1+\nu (t-1)\brace s}\,.
\label{snnut-2}
\ee

\section{Number of sites visited only once by the random walk}

\subsection{Discrete master equation for the joint probability distribution}

To get access to the statistics of the number $r$ of sites visited only once up to time $t$ we have first to evaluate the joint probability distribution $P_N(r,s,t)$, where $s\geq r$ has the same meaning as before. It satisfies the following master equation 
\be\fl
P_N(r,s,t)\!=\!\frac{s\!-\!r}{N}P_N(r,s,t\!-\!\!1)\!+\!\frac{r\!+\!1}{N}P_N(r\!+\!1,s,t\!-\!\!1)
\!+\!\frac{N\!\!-\!s\!+\!1}{N}P_N(r\!-\!1,s\!-\!\!1,t\!-\!\!1),
\label{pnrst-1}
\ee
with the boundary conditions $P_N(r,s<0,t)=P_N(r<0,s,t)=0$. The first (last) term on the right corresponds to a step towards one of the $s-r$ (respectively $N-s+1$) multi-visited (non-visited) sites at time $t-1$ and the middle term gives the contribution of a step towards one of the $r+1$ sites visited only once at time $t-1$.
The initial condition can be written as $P_N(r,s,0)=\delta_{r,0}\delta_{s,0}$, i.e., the walker is outside the lattice at $t=0$ and the walk starts at $t=1$ with equation~\eref{pnrst-1} giving $P_N(r,s,1)=\delta_{r,1}\delta_{s,1}$. The stationary solution is $P_N(r,s,t)=\delta_{r,0}\delta_{s,N}$

\subsection{Generating functions}
Let us introduce the multivariate generating function
\be\fl
\CP_N(x,y,z)=\sum_{t=0}^\infty\frac{z^t}{t!}\sum_{s=0}^\infty y^s\sum_{r=0}^\infty x^r P_N(r,s,t)
=1+\sum_{t=1}^\infty\frac{z^t}{t!}\sum_{s=0}^\infty y^s\sum_{r=0}^\infty x^r P_N(r,s,t)\,,
\label{pnxyz-1}
\ee
which is ordinary in $x,y$ and exponential in $z$. As shown in appendix~B, it satisfies the following partial differential equation:
\be
\CP_N(x,y,z)=\frac{x-1}{Nxy}\frac{\partial \CP_N}{\partial x}
+\frac{xy-1}{Nx}\frac{\partial \CP_N}{\partial y}
+\frac{1}{xy}\frac{\partial \CP_N}{\partial z}\,.
\label{pnxyz-2}
\ee
Let $\CP_N(x,y,z)=\CQ_N(\chi,\psi,\omega)$ where:
\be
\chi=\frac{yz(x-1)}{N}\,,\qquad \psi=1-y\,,\qquad \omega=y\rme^{z/N}\,.
\label{newvar}
\ee
With the new variables equation~\eref{pnxyz-2} transforms into:
\be\fl
\CQ_N(\chi,\psi,\omega)=\frac{1}{N}\left(\chi\frac{\partial\CQ_N}{\partial\chi}+\psi\frac{\partial\CQ_N}{\partial\psi}+\omega\frac{\partial\CQ_N}{\partial\omega}\right)+\frac{x-1}{Nx}\left(\frac{\partial\CQ_N}{\partial\chi}-\frac{\partial\CQ_N}{\partial\psi}\right)\,.
\label{qn-1}
\ee
The last term vanishes if $\chi$ and $\psi$ enter $\CQ_N$ under the combination $\chi+\psi$. Then $\CQ_N$ is homogeneous of degree $N$ in the new variables and thus takes the form 
\be
\CQ_N(\chi,\psi,\omega)=\left(\chi+\psi+\omega\right)^N\,,
\label{qn-2}
\ee
which translates into
\be
\CP_N(x,y,z)=\left[1+y\left(\rme^{z/N}-1+\frac{z(x-1)}{N}\right)\right]^N\,,
\label{pnxyz-3}
\ee
which indeed satisfies the initial and boundary conditions.

In the infinite lattice limit, one obtains:
\be
\CP_\infty(x,y,z)=\lim_{N\to\infty}\left(1+\frac{xyz}{N}\right)^N=\exp{xyz}\,.
\label{pinfxyz}
\ee
With $x=1$ one recovers $\CS_N(y,z)$ in equation~\eref{snyz}. The generating function for the probability distribution of the number of sites visited only once up to time $t$,
$R_N(r,t)=\sum_sP_N(r,s,t)$, is given by:
\be
\CR_N(x,z)=\CP_N(x,1,z)=\left[\rme^{z/N}+\frac{z(x-1)}{N}\right]^N\,.
\label{rnxz}
\ee

\subsection{Probability distributions}

\begin{figure}[!t]
\begin{center}
\includegraphics[width=7cm,angle=0]{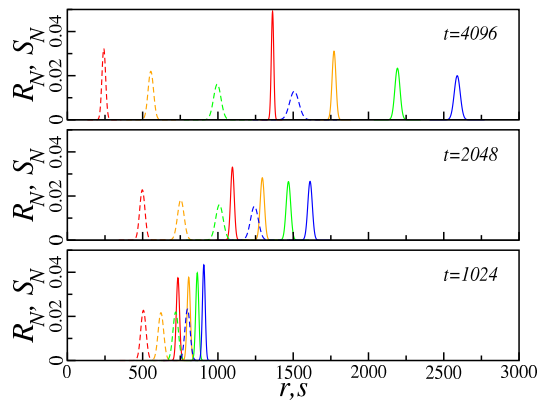}
\end{center}
\vglue -.5cm
\caption{Time evolution of the probability distributions $S_N(s,t)$ (full lines) and $R_N(r,t)$ (dashed lines) for different values of $N$  increasing from left to right (1448 (red), 2048 (orange), 2896 (green), 4096 (blue)).
\label{fig-2}
}
\end{figure}

An expansion of the generating function in equation~\eref{pnxyz-3} (see equation~\eref{egf-4} in appendix~C) leads to the following joint probability distribution~\footnote[2]{
When $r=0$ the sum in equation~\eref{pnrst-2} gives associated Stirling numbers of the second kind~\cite{oeis} 
$
T(t,s)=\sum_{k=0}^s(-1)^{k}{t\choose k}{t-k\brace s-k}\,.
$
}
:
\be
P_N(r,s,t)=
\frac{N^{\underline{s}}}{N^t}\sum_{k=r}^s(-1)^{k+r}{t\choose k}{k\choose r}{t-k\brace s-k}\,,
\qquad 0\leq r\leq s\,.
\label{pnrst-2}
\ee
According to equation~\eref{pinfxyz}, on the infinite lattice
\be
P_\infty(r,s,t)=\delta_{r,t}\delta_{s,t}
\label{pinfrst}
\ee
as expected, the walker visiting a new site at each step with probability one.

The probability distribution for the number of sites visited once up to time $t$ follows from~\eref{pnrst-2} by summing over $s$:
\be
R_N(r,t)=\frac{1}{N^t}\sum_{s=r}^t\sum_{k=r}^s(-1)^{k+r}{t\choose k}{k\choose r}{t-k\brace s-k}
N^{\underline{s}}\,.
\label{rnrt-1}
\ee
Changing the order of the sums, $\sum_{s=r}^t\sum_{k=r}^s\to\sum_{k=r}^t\sum_{s=k}^t$ and replacing the sum over $s$ by a sum over $j=s-k$ leads to:
\be
R_N(r,t)=\frac{1}{N^t}\sum_{k=r}^t(-1)^{k+r}{t\choose k}{k\choose r}\sum_{j=0}^{t-k}{t-k\brace j}
N^{\underline{j+k}}\,.
\label{rnrt-2}
\ee
Making use of the identities $N^{\underline{j+k}}=N^{\underline{k}}(N-k)^{\underline{j}}$ and $\sum_{j=0}^n{n\brace j}x^{\underline{j}}=x^n$ (see~\cite{graham94} p 262) one finally obtains:
\be
R_N(r,t)=\frac{1}{N^t}\sum_{k=r}^t(-1)^{k+r}{t\choose k}{k\choose r}(N-k)^{t-k}N^{\underline{k}}\,.
\label{rnrt-3}
\ee
The time evolution of $S_N(s,t)$ and $R_N(r,t)$ is shown in figure~\ref{fig-2} for different values of~$N$.

\section{Mean values, fluctuations and finite-size scaling behaviour}

\subsection{Moments of the probability distributions} 

\begin{figure}[!ht]
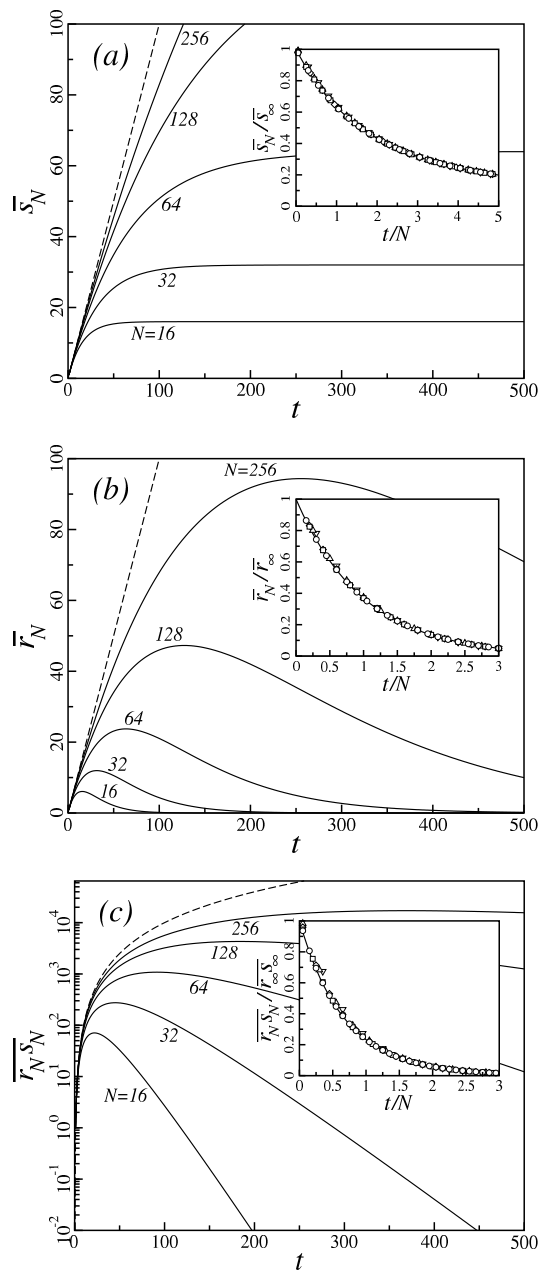

\begin{center}
\includegraphics[width=7cm,angle=0]{fig-3a.eps}
\vglue .3cm
\includegraphics[width=7cm,angle=0]{fig-3b.eps}
\vglue .3cm
\includegraphics[width=7cm,angle=0]{fig-3c.eps}
\end{center}
\vglue -.5cm
\caption{Time evolution for different lattice sizes of (a) $\overline{s_N(t)}$, the mean value of the number of distinct sites visited up to time $t$, (b) $\overline{r_N(t)}$, the mean value of the number of sites visited only once up to time $t$ and (c) $\overline{r_N(t)s_N(t)}$, the mean value of their product. The dashed lines correspond to the infinite system behaviour where $\overline{s_\infty(t)}=\overline{r_\infty(t)}=t$ and $\overline{r_\infty(t)s_\infty(t)}=t^2$, according to equation~\eref{pinfrst}. The insets show the data collapse obtained for the ratios of the mean values in finite and infinite systems as functions of the scaled time $w=t/N$ (see details in section~\ref{fss}). The scaling functions given in equation~\eref{fss-3} are indicated by full lines. Symbols correspond to finite-size results with $N=16$ (down triangle), 32 (up triangle), 64 (diamond), 128 (square), 256 (circle).
\label{fig-3}
}
\end{figure}

The mean number of distinct sites visited up to time $t$ is the coefficient of $z^t/t!$ in the $y$-derivative of the generating function~\eref{snyz} at~$y=1$~\footnote[3]{See appendix~D for direct calculations of the mean values}:
\be\fl
\overline{s_N(t)}=\left[\frac{z^t}{t!}\right]\left.\frac{\partial\CS_N(y,z)}{\partial y}\right|_{y=1}
=\left[\frac{z^t}{t!}\right]N\left(\rme^z-\rme^{\frac{N-1}{N}z}\right)
=N\left[1-\left(\frac{N-1}{N}\right)^t\right]\,.
\label{snt}
\ee
As in the infinite system the initial growth is linear. The approach to the saturation value, $N$, is exponential with a relaxation time $t_{\rm r}$ such that $\lim_{N\to\infty}t_{\rm r}=N$ (see figure~\ref{fig-3}(a)).

The mean number of sites visited once up to time $t$ is deduced in the same way from $\CR_N(x,z)$ in equation~\eref{rnxz}
\be\fl
\overline{r_N(t)}=\left[\frac{z^t}{t!}\right]\left.\frac{\partial\CR_N(x,z)}{\partial x}\right|_{x=1}
=\left[\frac{z^t}{t!}\right]N\left(z\,\rme^{\frac{N-1}{N}z}\right)
=t\left(\frac{N-1}{N}\right)^{t-1}\,.
\label{rnt}
\ee
It grows linearly at short time as $\overline{s_N(t)}$, goes through a maximum at $t_{\rm max}\approx N$ and then decays exponentially with a relaxation time $t_{\rm r}=N$, asymptotically (see figure~\ref{fig-3}(b)).

The second moments are given by
\bea
\fl\overline{s^2_N(t)}&=\left[\frac{z^t}{t!}\right]\left.\frac{\partial}{\partial y}\left[y\,\frac{\partial\CS_N(y,z)}{\partial y}\right]\right|_{y=1}
=\overline{s_N(t)}+\left[\frac{z^t}{t!}\right]
N(N-1)\left(\rme^z-2\rme^{\frac{N-1}{N}z}+\rme^{\frac{N-2}{N}z}\right)\nonumber\\
\fl&=N^2-N(2N-1)\left(\frac{N-1}{N}\right)^t+N(N-1)\left(\frac{N-2}{N}\right)^t\,,
\label{sn2t}
\eea
and 
\bea
\fl\overline{r^2_N(t)}&=\left[\frac{z^t}{t!}\right]\left.\frac{\partial}{\partial x}\left[x\,\frac{\partial\CR_N(x,z)}{\partial x}\right]\right|_{x=1}
=\overline{r_N(t)}+\left[\frac{z^t}{t!}\right]
\frac{N-1}{N}z^2\rme^{\frac{N-2}{N}z}\,,\nonumber\\
\fl&=t\left(\frac{N-1}{N}\right)^{t-1}\!\!\!
+t(t-1)\frac{N-1}{N}\left(\frac{N-2}{N}\right)^{t-2}\,.
\label{rn2t}
\eea
The mean value of the product of the two variables is obtained by taking a second derivative of $\CP_N(x,y,z)$ in equation~\eref{pnxyz-3}
\bea
\fl\overline{r_N(t)s_N(t)}&=\left[\frac{z^t}{t!}\right]
\left.\frac{\partial^2\CP_N(x,y,z)}{\partial x\partial y}\right|_{x=y=1}
=\left[\frac{z^t}{t!}\right]\left[Nz\rme^{\frac{N-1}{N}z}-(N-1)z\rme^{\frac{N-2}{N}z}\right]\nonumber\\
\fl&=Nt\left(\frac{N-1}{N}\right)^{t-1}-(N-1)t\left(\frac{N-2}{N}\right)^{t-1}\,.
\label{rnsnt}
\eea
It behaves as $N\overline{r_N(t)}$ when $t\gg N$ (see figure~\ref{fig-3}(c)).

\subsection{Variances and covariance}

\begin{figure}[!t]
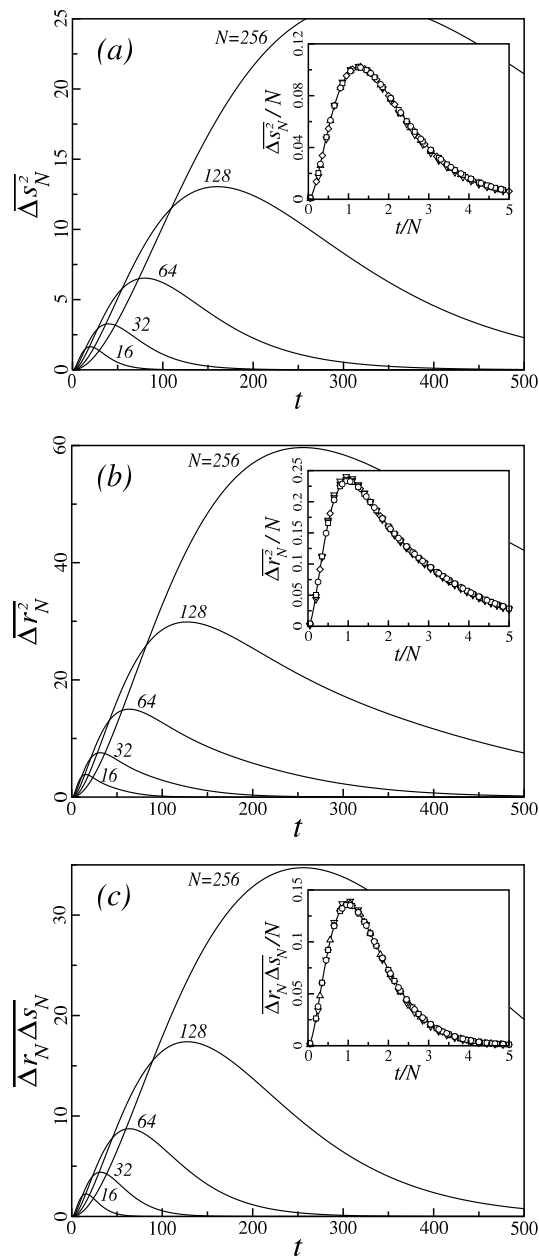

\begin{center}
\includegraphics[width=7cm,angle=0]{fig-4a.eps}
\vglue .3cm
\includegraphics[width=7cm,angle=0]{fig-4b.eps}
\vglue .3cm
\includegraphics[width=7cm,angle=0]{fig-4c.eps}
\end{center}
\vglue -.5cm
\caption{Time evolution for different lattice sizes of (a) the variance $\overline{\Delta s_N^2(t)}$ of the number of distinct sites visited up to time $t$, (b) the variance $\overline{\Delta r_N^2(t)}$ of the number of sites visited only once up to time $t$ and (c) the covariance $\overline{\Delta r_N(t)\Delta s_N(t)}$. The insets show the data collapse obtained with the scaled variables defined in section~\ref{fss}. The scaling functions given in equation~\eref{fss-5} are indicated by full lines and the symbols have the same meaning as in figure~\ref{fig-3}.
\label{fig-4}
}
\end{figure}

Combining these results, one obtains the variances (figures~\ref{fig-4}(a) and~(b))
\be\fl
\overline{\Delta s^2_N(t)}=\overline{s^2_N(t)}-\overline{s_N^2(t)}
=N\left(\frac{N-1}{N}\right)^t\!\!+N(N-1)\left(\frac{N-2}{N}\right)^t\!\!-N^2\left(\frac{N-1}{N}\right)^{2t}\!\!\!\!,
\label{dsn2t}
\ee

\be\fl
\overline{\Delta r^2_N(t)}=t\left(\frac{N-1}{N}\right)^{t-1}\!\!\!\!\!+t(t-1)\frac{N-1}{N}\left(\frac{N-2}{N}\right)^{t-2}\!\!\!\!\!-t^2\left(\frac{N-1}{N}\right)^{2t-2}\!\!\!\!\!\!\!.
\label{drn2t}
\ee
and the covariance (figure~\ref{fig-4}(c))
\be\fl
\overline{\Delta r_N(t)\Delta s_N(t)}=\overline{r_Ns_N(t)}-\overline{r_N(t)}\,\overline{s_N(t)}
=Nt\left(\frac{N-1}{N}\right)^{2t-1}\!\!\!\!\!-(N-1)t\left(\frac{N-2}{N}\right)^{t-1}\!\!\!\!\!.
\label{drnsnt}
\ee
The three functions have a similar time-dependence. The initial growth is quadratic in $t$. The fluctuations are maximum at values $t_{\rm max}$ close to $N$ and the decay is exponential at long time.

\subsection{Finite-size scaling}
\label{fss}
In order to determine how finite-size effects alter the properties of the fully-connected lattice and what are the appropriate scaling variables, we first examine the case of a periodic lattice in $d$ dimensions and then extend the results to $d$ infinite.

Let us first consider the scaling behaviour of $\overline{s_\infty(t)}$, the mean number of distinct sites visited up to time $t$ by a random walk on the infinite lattice in $d$ dimensions. The fractal dimension of the walk is $d_f=2$ such that the typical size of the walk at time $t$ is $R(t)\sim t^{1/d_f}$. Below the critical dimension $d_{\rm c}=2$, the fractal dimension of the walk is larger than the Euclidean dimension $d$ and the exploration of space is compact, thus  
$\overline{s_\infty(t)}\sim R^d(t)\sim t^{d/d_f}$. Above $d_{\rm c}$, multiple visits are irrelevant and, to leading order, $\overline{s_\infty(t)}\sim t$. This new regime sets in with a logarithmic correction at $d_{\rm c}$ (see equation~\eref{sinft}).
For $\overline{r_\infty(t)}$, due to the irrelevance of multiple visits, the same behaviour as for $\overline{s_\infty(t)}$ is obtained above $d_{\rm c}$~\cite{montroll65}. In contrast, the behaviour differs at and below $d_{\rm c}$ as indicated in equation~\eref{rinft}.

On a finite system with size $L$ and $N=L^d$ sites, finite-size effects are governed by the dimensionless ratio $R(t)/L$ or $t^{d/d_f}/N$ and the finite-size scaling Ansatz, valid when $N\gg1$ and $t\gg1$, takes the following form~\cite{fisher72,hamer80}:
\be
G_N(t)=G_\infty(t)\,\phi_G\left(\frac{t^{d/d_f}}{N}\right)\,,\qquad \lim_{w\to 0}\phi_G(w)=1\,.
\label{fss-1}
\ee
This finite-size behaviour applies only below $d_{\rm c}$. Above $d_{\rm c}$ and, in particular, on the fully-connected lattice for which $d=\infty$, the scaling relations are verified with $d$ replaced by $d_{\rm c}$, the upper critical dimension above which the critical exponents remain constant~\cite{botet82,botet83}. Thus we have:
\be\fl
\overline{s_N(t)}\!=\!\overline{s_\infty(t)}\,\phi_s(t/N),\,\,\overline{r_N(t)}\!=\!\overline{r_\infty(t)}\,\phi_r(t/N),\,\, \overline{r_N(t)s_N(t)}\!=\!\overline{r_\infty(t)s_\infty(t)}\,\phi_{rs}(t/N).
\label{fss-2}
\ee
Since above $d_{\rm c}$  $\overline{s_\infty(t)}\sim t$ and $\lim_{t\to\infty}\overline{s_N(t)}=N$, the time-dependence disappears in this limit only if the scaling function behaves as $w^{-1}$ with a scaled variable $w=t/N$.

On the fully-connected lattice, according to~\eref{pinfrst}, $\overline{s_\infty(t)}=\overline{r_\infty(t)}=t$ 
and $\overline{r_\infty(t)s_\infty(t)}=t^2$. Thus equations~\eref{snt}, \eref{rnt} and~\eref{rnsnt} lead to the scaling functions~\footnote[4]{The scaling functions are obtained by making use of the identity $(1-a/N)^{bt-c}=\rme^{-abw}\left[1+(c-abw/2)a/N+O(N^{-2})\right]$, where $w=t/N$ is the scaled time.}:
\be
\phi_s(w)=\frac{1-\rme^{-w}}{w}\,,\quad \phi_r(w)=\rme^{-w}\,,
\quad \phi_{rs}(w)=\frac{\rme^{-w}-\rme^{-2w}}{w}\,.
\label{fss-3}
\ee

As shown in the insets of figures~\ref{fig-3}(a)--(c), a good data collapse is obtained in agreement with these scaling functions for the whole range of values of $w=t/N$, provided that $N$ and $t$ are not too small. 

A different normalization is required for the variances which vanish when $N\to\infty$ according to equation~\eref{pinfrst}. Since both the variances in equations \eref{dsn2t} and~\eref{drn2t} and the covariance in  equation~\eref{drnsnt} scale as $N$, the following scaling forms are appropriate:
\be\fl
\overline{\Delta s^2_N(t)}=N\,\tau_s(t/N)\,,\quad \overline{\Delta r^2_N(t)}=N\,\tau_r(t/N)\,,\quad \overline{\Delta r_N(t)\Delta s_N(t)}=N\,\tau_{rs}(t/N)\,.
\label{fss-4}
\ee
From equations~\eref{dsn2t}, \eref{drn2t} and~\eref{drnsnt} one deduces the following scaling functions:
\be\fl
\tau_s(w)\!=\!\rme^{-w}\!-\!(w\!+\!1)\rme^{-2w}\!,\quad \tau_r(w)\!=w\!\left[\rme^{-w}\!\!-(w^2\!\!-\!w\!+1\!)\rme^{-2w}\right]\!,\quad \tau_{rs}(w)\!=\!w^2\rme^{-2w}\!.
\label{fss-5}
\ee
Here too the data collapse is quite good, even for relatively small values of $N$ (see the insets 
in figures~\ref{fig-4}(a)--(c)). The maxima of the scaling functions are located at 
$w_{\rm max}=1$ for $\overline{\Delta r^2_N}$ and $\overline{\Delta r_N\Delta s_N}$ and at 
$w_{\rm max}=1.256431\dots$, the positive solution of $w=\ln(1+2w)$, for $\overline{\Delta s^2_N}$.

\section{Probability densities in the scaling limit}

\begin{figure}[!t]
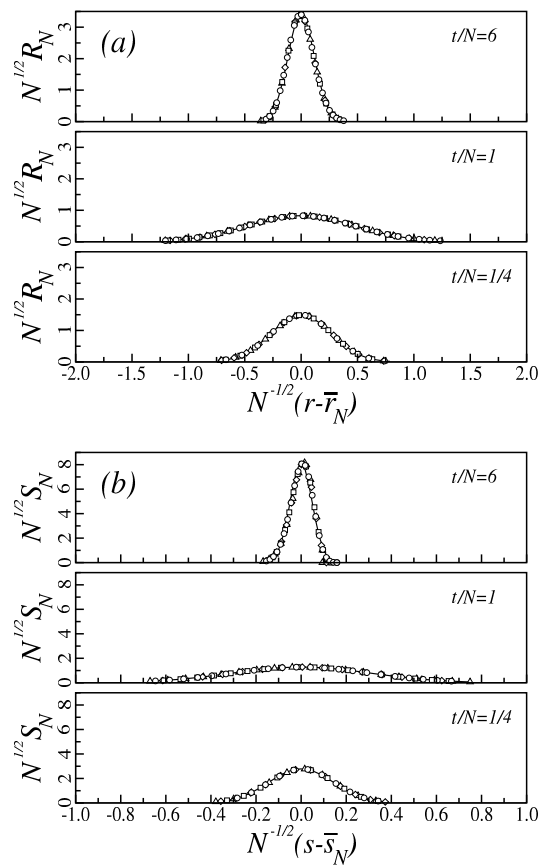

\begin{center}
\includegraphics[width=7cm,angle=0]{fig-5a.eps}
\vglue .3cm
\includegraphics[width=7cm,angle=0]{fig-5b.eps}
\end{center}
\vglue -.5cm
\caption{Data collapse obtained for the scaled probability distributions at different values of the scaled time $w=t/N$: 
(a) $N^{1/2}R_N(r,t)$ as a function of $u=N^{-1/2}(r-\overline{r_N(t)})$ and (b) $N^{1/2}S_N(r,t)$ as a function of $v=N^{-1/2}(s-\overline{s_N(t)})$. The different lattice sizes are $N=1448$ (up triangle), $2048$ (diamond), $2896$ (square), $4096$ (circle)). The Gaussian densities obtained in the scaling limit, (a) $\rho(u,w)$ and (b) $\sigma(v,w)$, given by equation~\eref{gauss}, are indicated by full lines.
\label{fig-5}
}
\end{figure}

In this section we work in the scaling limit ($t\to\infty$, $N\to\infty$, $w=t/N$ fixed) and we introduce the scaling variables
\be
u=\frac{r-\overline{r_N(t)}}{N^{1/2}}\,,\qquad v=\frac{s-\overline{s_N(t)}}{N^{1/2}}\,,
\label{scalvar}
\ee
the form of which follows from equation~\eref{fss-4}.

In this continuum limit the probability distributions $R_N(r,t)$, $S_N(s,t)$ and $P_N(r,s,t)$ transform into the probability densities $\rho(u,w)$, $\sigma(v,w)$ and $\Pi(u,v,w)$, respectively. Since $\rmd u=N^{-1/2}\rmd r$ and 
$\rmd v=N^{-1/2}\rmd s$, the conservation of probability leads to the following relations in the scaling limit: 
$\rho=N^{1/2}R_N$, $\sigma=N^{1/2}S_N$ and $\Pi=NP_N$.

The scaling behaviour of the probability densities is illustrated in figures \ref{fig-5}(a) and~(b) for $\rho$ and $\sigma$. These figures strongly suggest a Gaussian behaviour for both densities. In order to verify this point we rewrite the master equation \eref{pnrst-1} in terms of the scaled variables and keep the leading contributions in an expansion in powers of $N^{-1/2}$. The calculation is lengthy but straightforward (details are given in~\ref{pdeuvw}) and leads to the following partial differential equation:
\bea
\frac{\partial\Pi}{\partial w}&=\frac{\rme^{-w}}{2}\left[w+1-(w-1)^2\rme^{-w}\right]
\frac{\partial^2\Pi}{\partial u^2}+\frac{\rme^{-w}}{2}\left(1-\rme^{-w}\right)
\frac{\partial^2\Pi}{\partial v^2}\nonumber\\
&+\rme^{-w}\left[1+(w-1)\rme^{-w}\right]\frac{\partial^2\Pi}{\partial u\partial v}
+(u+v)\frac{\partial \Pi}{\partial u}+v\frac{\partial \Pi}{\partial v} +2\Pi\,.
\label{piw}
\eea
Using Maple$\rm\sp{TM}$ we verified that the bivariate Gaussian density
\bea
\Pi(u,v,w)&=&\frac{1}{2\pi\sqrt{\Delta(w)}}
\exp\left[-\frac{\tau_s(w)u^2-2\tau_{rs}(w)uv+\tau_r(w)v^2}{2\Delta(w)}\right]\,,\nonumber\\
\ \ \ \ \ \ \Delta(w)&=&\tau_r(w)\tau_s(w)-\tau_{rs}^2(w)\,,
\label{bivgauss}
\eea
is indeed solution of equation~\eref{piw} with $\tau_r$, $\tau_s$ and $\tau_{rs}$ given by equation~\eref{fss-5}.
An integration over either $v$ or $u$ leads to the Gaussian probability densities
\be\fl
\rho(u,w)=\frac{1}{\sqrt{2\pi\tau_r(w)}}
\exp\left[-\frac{u^2}{2\tau_{r}(w)}\right]\,,\qquad
\sigma(v,w)=\frac{1}{\sqrt{2\pi\tau_s(w)}}
\exp\left[-\frac{v^2}{2\tau_{s}(w)}\right]\,.
\label{gauss}
\ee
The evolution of these probability densities is illustrated in figures~\ref{fig-5}(a) and~(b) for three values of the scaled time $w=t/N$.

In the case of $\nu$ walkers, when $N$ and $\nu t$ are large, the probability distribution for the number of distinct sites visited up to time $t$ in equation \eref{snnut-1} leads to the same Gaussian density 
$\sigma_\nu(v,w)=\sigma(v,w)$ with $v=[s-\overline{s_N(\nu t)}]/N^{1/2}$ and $w=\nu t/N$.

\section{Conclusion}
In this work we have studied the statistics of the number of sites visited by a random walk up to time $t$ on a fully-connected lattice with $N$ sites. Exact expressions have been obtained for the probability distributions  
$S_N(s,t)$ and $R_N(r,t)$ associated with the total number $s$ of distinct sites visited and the number $r$ of sites visited once. This last distribution was deduced from the joint probability distribution, $P_N(r,s,t)$, itself derived via generating function techniques. The mean values, variances and covariance of $r$ and $s$ have been calculated by taking derivatives of the appropriate generating functions. Their finite-size scaling behaviour have been determined, allowing us to define centered and scaled variables $u$ and  $v$, corresponding respectively to $r$ and $s$, and a scaled time $w=t/N$. Using these new variables, a partial differential equation for the joint probability density $\Pi(u,v,w)$ have been obtained in the scaling limit. The solution is a bivariate Gaussian density thus the scaled variables $u$ and $v$ both display Gaussian fluctuations. 

We believe our results are representative of the behaviour on periodic lattices above~$d_{\rm c}$. The same type of finite-size scaling should be observed for the number of distinct sites visited for $d\geq3$. A different behaviour is expected for $d\leq2$, involving logarithmic corrections at $d=2$ and, as explained in section 4.3, the scaling variable $t^{d/d_f}/N$ with $N=L^d$ for $d<2$. Thus a finite-size scaling study of the mean values and the probability distributions of $r$ and $s$ in $1d$ and $2d$ would be of interest. 

In the case of $\nu$ walkers, the finite-size effects on the number of common sites visited deserves also some attention. 

\appendix

\section{Generating function for $S_N(s,t)$}
\label{gfs}
\setcounter{section}{1}
It is convenient to define a bivariate generating function 
\be 
\CS_N(y,z)=\sum_{s=0}^\infty y^s\sum_{t=0}^\infty\frac{z^t}{t!}\,S_N(s,t)\,,
\label{gfs-1}
\ee
which is ordinary in $y$ and exponential in $z$. Inserting the probability distribution under the form given in equation~\eref{snst-2}, one obtains
\be
\CS_N(y,z)=\left.\sum_{s=0}^N{N\choose s}(y\BD)^s\sum_{t=0}^\infty\frac{(z\eta/N)^t}{t!}\right|_{\eta=0}
\!\!\!\!=\left.(\BI+y\BD)^N\rme^{z\eta/N}\right|_{\eta=0}\,,
\label{gfs-2}
\ee
where $\BI$ is the identity operator and $\BD$ is the finite-difference operator acting on $\eta$. Since
\be\fl
(\BI+y\BD)\rme^{z\eta/N}=\rme^{z\eta/N}+y\left[\rme^{z(\eta+1)/N}-\rme^{z\eta/N}\right]
=\left[1+y\left(\rme^{z/N}-1\right)\right]\rme^{z\eta/N}\,,
\label{gfs-3}
\ee
repeating this operation $N$ times leads to
\be
\CS_N(y,z)=\left.\left[1+y\left(\rme^{z/N}-1\right)\right]^N\rme^{z\eta/N}\right|_{\eta=0}
=\left[1+y\left(\rme^{z/N}-1\right)\right]^N\,.
\label{gfs-4}
\ee

\section{Partial differential equation for $\CP_N(x,y,z)$}
\label{pdexyz}
Replacing $P_N(r,s,t)$ on the right-hand side of equation~\eref{pnxyz-1} by its expression in~\eref{pnrst-1} and using the boundary conditions,
one has:
\bea
\fl\CP_N(x,y,z)\!=\!1\!+\!\!\frac{1}{N}\!\sum_{t=1}^\infty\!\frac{z^t}{t!}\!
\sum_{s=0}^\infty\! sy^s\sum_{r=0}^\infty\! x^r P_N(r,s,t\!-\!1)
\!-\!\frac{1}{N}\!\sum_{t=1}^\infty\!\frac{z^t}{t!}\!\sum_{s=0}^\infty\! y^s
\sum_{r=0}^\infty\! rx^rP_N(r,s,t\!-\!1)\nonumber\\
\!\!\!\!\!\!\!\!\!\!\!\!\!\!\!\!\!\!\!\!\!\!\!+\!\frac{1}{N}\!\sum_{t=1}^\infty\!\frac{z^t}{t!}\!\sum_{s=0}^\infty\! y^s
\sum_{r=0}^\infty\! (r\!+\!1)x^rP_N(r\!+\!1,s,t\!-\!1)
\!+\!\sum_{t=1}^\infty\!\frac{z^t}{t!}\!\sum_{s=1}^\infty\! y^s
\sum_{r=1}^\infty\!x^rP_N(r\!-\!1,s\!-\!1,t\!-\!1)\nonumber\\
\!\!\!\!\!\!\!\!\!\!\!\!\!\!\!\!\!\!\!\!\!\!\!-\!\frac{1}{N}\!\sum_{t=1}^\infty\!\frac{z^t}{t!}\!\sum_{s=1}^\infty\!(s\!-\!1) y^s
\sum_{r=1}^\infty\!x^rP_N(r\!-\!1,s\!-\!1,t\!-\!1)\,.
\label{pde-1}
\eea
Using the changes of variables $t-1\to t$ and $s-1\to s$, $r\pm 1\to r$ when appropriate, one obtains:
\bea
\fl\CP_N(x,y,z)\!=\!1\!+\!\!\frac{y}{N}\!\sum_{t=0}^\infty\!\frac{z^{t+1}}{(t\!+\!\!1)!}\!
\sum_{s=0}^\infty\!\! sy^{s-1}\!\sum_{r=0}^\infty\! \!x^r\! P_N(r,s,t)
\!-\!\frac{x}{N}\!\sum_{t=0}^\infty\!\frac{z^{t+1}}{(t\!+\!\!1)!}\!\sum_{s=0}^\infty\! y^s\!\sum_{r=0}^\infty\!\! rx^{r-\!1}\!P_N(r,s,t)\nonumber\\
\!\!\!\!\!\!\!\!\!\!\!\!\!\!\!\!+\frac{1}{N}\!\sum_{t=0}^\infty\!\frac{z^{t+1}}{(t+1)!}\!\sum_{s=0}^\infty\! y^s
\sum_{r=0}^\infty\! rx^{r-1}P_N(r,s,t)
\!+\!xy\sum_{t=0}^\infty\!\frac{z^{t+1}}{(t+1)!}\!\sum_{s=1}^\infty\! y^s
\sum_{r=0}^\infty\!x^rP_N(r,s,t))\nonumber\\
\!\!\!\!\!\!\!\!\!\!\!\!\!\!\!\!-\frac{xy^2}{N}\!\sum_{t=0}^\infty\!\frac{z^{t+1}}{(t+1)!}\!\sum_{s=0}^\infty\!sy^{s-1}\sum_{r=0}^\infty\!x^rP_N(r,s,t)\,.
\label{pde-2}
\eea
Taking the partial derivative of both sides with respect to $z$ leads to:
\be
\frac{\partial\CP_N}{\partial z}=\frac{y(1-xy)}{N}\frac{\partial\CP_N}{\partial y}
+\frac{1-x}{N}\frac{\partial\CP_N}{\partial x}+xy\CP_N(x,y,z)\,,
\label{pde-3}
\ee
from which equation~\eref{pnxyz-2} follows.

\section{Expansion of the generating function}
\label{expans}

Expanding~\eref{pnxyz-3}, one obtains:
\bea
\CP_N(x,y,z)&=&\sum_{s=0}^Ny^s{N\choose s}(\rme^{z/N}-1)^s\left[1+\frac{z(x-1)}{N(\rme^{z/N}-1)}\right]^s\nonumber\\
&=&\sum_{s=0}^\infty y^sN^{\underline{s}}\sum_{k=0}^s\left(\frac{z}{N}\right)^k\frac{(x-1)^k}{k!}
\frac{(\rme^{z/N}-1)^{s-k}}{(s-k)!}\,.
\label{egf-1}
\eea
Making use of the exponential generating function of the Stirling numbers of the second kind (\cite{graham94} p 351)
\be
\frac{(\rme^u-1)^m}{m!}=\sum_{n=m}^\infty{n\brace m}\frac{u^n}{n!}\,,
\label{egfs}
\ee
with $u=z/N$, $m=s-k$ and $n=t-k$, \eref{egf-1} can be rewritten as:
\be
\CP_N(x,y,z)=\sum_{s=0}^\infty y^sN^{\underline{s}}\sum_{k=0}^s\frac{(x-1)^k}{k!}
\sum_{t=0}^\infty\frac{(z/N)^t}{(t-k)!}{t-k\brace s-k}\,.
\label{egf-2}
\ee
Note that the last sum actually starts at $t=s$ since the Stirling numbers vanish when $t<s$. Changing the order of the sums leads to:
\bea
\CP_N(x,y,z)&=&\sum_{t=0}^\infty\frac{z^t}{t!}\sum_{s=0}^\infty y^s\frac{N^{\underline{s}}}{N^t}
\sum_{k=0}^s{t\choose k}{t-k\brace s-k}(x-1)^k\nonumber\\
&=&\sum_{t=0}^\infty\frac{z^t}{t!}\sum_{s=0}^\infty y^s\frac{N^{\underline{s}}}{N^t}
\sum_{k=0}^s{t\choose k}{t-k\brace s-k}\sum_{r=0}^kx^r(-1)^{k+r}{k\choose r}\,.
\label{egf-3}
\eea
Changing $\sum_{k=0}^s\sum_{r=0}^k$ into $\sum_{r=0}^s\sum_{k=r}^s$ gives
\be
\CP_N(x,y,z)=\sum_{t=0}^\infty\frac{z^t}{t!}\sum_{s=0}^\infty y^s
\sum_{r=0}^sx^r\frac{N^{\underline{s}}}{N^t}\sum_{k=r}^s(-1)^{k+r}{t\choose k}{k\choose r}{t-k\brace s-k}\,,
\label{egf-4}
\ee
from which, comparing to equation~\eref{pnxyz-1}, one extracts the probability distribution function
$P_N(r,s,t)$ in equation~\eref{pnrst-2}.

When $x=1$ in the first line of~\eref{egf-3}, i.e. for $\CS_N(y,z)$, the term $k=0$ alone contributes to the last sum and, comparing to \eref{snyz}, one recovers the final expression of $S_N(s,t)$ in equation~\eref{snst-1}.

\section{Direct calculation of some mean values}
\label{meanval}

The mean values $\overline{s_N(t)}$ in equation \eref{snt} and $\overline{r_N(t)}$ in equation \eref{rnt} can be obtained directly as follows. Let $n_i^{(k)}(t)$ be a binary indicator associated with site $i$ such that 
$n_i^{(k)}(t)=1$ when this site has been visited $k$ times by the walker up to time $t$ and $n_i^{(k)}(t)=0$ otherwise. Then the number of sites visited $k$ times for a given realization of the walk is 
$s_N^{(k)}(t)=\sum_{i=1}^Nn_i^{(k)}(t)$ and its mean value is given by
\be
\overline{s_N^{(k)}(t)}=\sum_{i=1}^N\overline{n_i^{(k)}(t)}=Np_k(t)\,,
\label{mv-1}
\ee
where
\be
p_k(t)={t\choose k}\left(\frac{1}{N}\right)^k\left(\frac{N-1}{N}\right)^{t-k}
\label{mv-2}
\ee
is the probability to visit the same site exactly $k$ times in $t$ steps. Thus one obtains
\be
\overline{s_N(t)}=N-\overline{s_N^{(0)}(t)}=N\left[1-\left(\frac{N-1}{N}\right)^t\right]\,,
\label{mv-3}
\ee
in agreement with equation~\eref{snt} and
\be
\overline{r_N(t)}=\overline{s_N^{(1)}(t)}=t\left(\frac{N-1}{N}\right)^{t-1}\,,
\label{mv-4}
\ee
in agreement with equation~\eref{rnt}.

In the same way
\be
\overline{s_N^{(k)}(t)s_N^{(l)}(t)}=\sum_{i,j=1}^N\overline{n_i^{(k)}(t)n_j^{(l)}(t)}=N(N-1)p_{kl}(t)+Np_k(t)\delta_{k,l}\,,
\label{mv-5}
\ee
where $p_k(t)$ is given by equation~\eref{mv-2} and
\be
p_{kl}(t)={t\choose k}{t-k\choose l}\left(\frac{1}{N}\right)^{k+l}\left(\frac{N-2}{N}\right)^{t-k-l}
\label{mv-6}
\ee
is the probability that two given sites have been respectively visited $k$ and $l$ times, up to time $t$.
Thus, making use of equations~\eref{mv-5}, \eref{mv-6} and \eref{mv-2}, one obtains
\bea
\overline{s_N^2(t)}&=\overline{\left[N-s_N^{(0)}(t)\right]^2}
=N^2-2N\overline{s_N^{(0)}(t)}+\overline{\left[s_N^{(0)}(t)\right]^2}\nonumber\\
&=N^2-N(2N-1)\left(\frac{N-1}{N}\right)^t+N(N-1)\left(\frac{N-2}{N}\right)^t\,,
\label{mv-7}
\eea
in agreement with equation~\eref{sn2t} and
\be
\overline{r_N^2(t)}=\overline{\left[s_N^{(1)}(t)\right]^2}=t\left(\frac{N-1}{N}\right)^{t-1}\!\!\!
+t(t-1)\frac{N-1}{N}\left(\frac{N-2}{N}\right)^{t-2}\,.
\label{mv-8}
\ee
in agreement with equation~\eref{rn2t}.

\section{Partial differential equation for $\Pi(u,v,w)$}
\label{pdeuvw}

Making use of equations~\eref{fss-2} and~\eref{fss-3}, the scaled variables in equation~\eref{scalvar} 
are given by 
\be
u=-w\rme^{-w}N^{1/2}+\frac{r}{N^{1/2}}\,,\quad v=\left(\rme^{-w}-1\right)N^{1/2}+\frac{s}{N^{1/2}}\,,\quad
w=\frac{t}{N}\,,
\label{scalvar-1}
\ee
with the following partial derivatives:
\be\fl
\frac{\partial u}{\partial r}=\frac{1}{N^{1/2}}\,,\quad 
\frac{\partial u}{\partial t}=(w-1)\frac{\rme^{-w}}{N^{1/2}}\,,\quad
\frac{\partial v}{\partial s}=\frac{1}{N^{1/2}}\,,\quad 
\frac{\partial v}{\partial t}=-\frac{\rme^{-w}}{N^{1/2}}\,,\quad
\frac{\partial w}{\partial t}=\frac{1}{N}\,.
\label{der-1}
\ee
Other derivatives either vanish or are of order $N^{-3/2}$ or higher.

To obtain a partial differential equation for $\Pi(u,v,w)$, one first multiplies the master equation~\eref{pnrst-1} by $N$ since $NP_N(r,s,t)$ gives $\Pi[u(r,t),v(s,t),w(t)]$ in the scaling limit. Using equation~\eref{scalvar-1} to rewrite the prefactors one obtains:
\bea
\fl\Pi\!=\!\left(1\!-\!\rme^{-w}\!-\!w\rme^{-w}\!-\frac{v\!-\!u}{N^{1/2}}\right)\!NP_N(r,s,t\!-\!1)
+\left(w\rme^{-w}\!+\!\frac{u}{N^{1/2}}\!+\!\frac{1}{N}\right)\!NP_N(r\!+\!1,s,t\!-\!1)\nonumber\\
+\left(\rme^{-w}\!-\!\frac{v}{N^{1/2}}\!+\!\frac{1}{N}\right)\!NP_N(r\!-\!1,s-1,t\!-\!1)\,.
\label{pi}
\eea  
In the next step, one expands $P_N$ on the right-hand side. To keep terms of order at most $N^{-1}$ in the final equation, the expansion of $P_N$ is up to terms of second order:
\bea
\fl P_N(r,s,t\!-\!1)\!=\!P_N(r,s,t)-\frac{\partial P_N}{\partial t}+\frac{1}{2}\frac{\partial^2\! P_N}{\partial t^2}\,,\nonumber\\
\fl P_N(r\!+\!1,s,t\!-\!1)\!=\!P_N(r,s,t)+\frac{\partial P_N}{\partial r}-\frac{\partial P_N}{\partial t}+\frac{1}{2}\frac{\partial^2\! P_N}{\partial r^2}-\frac{\partial^2\! P_N}{\partial r\partial t}+\frac{1}{2}\frac{\partial^2\! P_N}{\partial t^2}\,,\nonumber\\
\fl P_N(r\!-\!1,s\!-\!1,t\!-\!1)\!=\!P_N(r,s,t)-\frac{\partial P_N}{\partial r}-\frac{\partial P_N}{\partial s}
-\frac{\partial P_N}{\partial t}+\frac{1}{2}\frac{\partial^2\! P_N}{\partial r^2}+\frac{1}{2}\frac{\partial^2\! P_N}{\partial s^2}+\frac{1}{2}\frac{\partial^2\! P_N}{\partial t^2}\nonumber\\
\ \ \ \ \ \ \ \ \ \ \ \ \ \ \ \ \ \ \ \ \ \ \ \ +\frac{\partial^2\! P_N}{\partial r\partial s}
+\frac{\partial^2\! P_N}{\partial r\partial t}+\frac{\partial^2\! P_N}{\partial s\partial t}\,.
\label{pnexpans}
\eea
Thus the following derivatives are needed:
\bea
\fl N\frac{\partial P_N}{\partial r}\!=\!\frac{1}{N^{1/2}}\frac{\partial\Pi}{\partial u},\quad
 N\frac{\partial P_N}{\partial s}\!=\!\frac{1}{N^{1/2}}\frac{\partial\Pi}{\partial v},\quad
N\frac{\partial P_N}{\partial t}\!=\!(w\!-\!1)\frac{\rme^{-w}}{N^{1/2}}\frac{\partial\Pi}{\partial u}
\!-\!\frac{\rme^{-w}}{N^{1/2}}\frac{\partial\Pi}{\partial v}\!
+\!\frac{1}{N}\frac{\partial\Pi}{\partial w},\nonumber\\
\fl N\frac{\partial^2\! P_N}{\partial r^2}=\frac{1}{N}\frac{\partial^2\Pi}{\partial u^2}\,,\quad
N\frac{\partial^2\! P_N}{\partial s^2}=\frac{1}{N}\frac{\partial^2\Pi}{\partial v^2}\,,\quad
N\frac{\partial^2\! P_N}{\partial r\partial s}=\frac{1}{N}\frac{\partial^2\Pi}{\partial u\partial v}\,,\nonumber\\
\fl N\frac{\partial^2\! P_N}{\partial t^2}=(w-1)^2\frac{\rme^{-2w}}{N}\frac{\partial^2\Pi}{\partial u^2}
-2(w-1)\frac{\rme^{-2w}}{N}\frac{\partial^2\Pi}{\partial u\partial v}
+\frac{\rme^{-2w}}{N}\frac{\partial^2\Pi}{\partial v^2}\,,\nonumber\\
\fl N\frac{\partial^2\! P_N}{\partial r\partial t}\!=
\!(w\!-\!1)\frac{\rme^{-w}}{N}\frac{\partial^2\Pi}{\partial u^2}
\!-\!\frac{\rme^{-w}}{N}\frac{\partial^2\Pi}{\partial u\partial v},\quad
N\frac{\partial^2\! P_N}{\partial s\partial t}\!=\!-\frac{\rme^{-w}}{N}\frac{\partial^2\Pi}{\partial v^2}
\!+\!(w\!-\!1)\frac{\rme^{-w}}{N}\frac{\partial^2\Pi}{\partial u\partial v}.
\label{der-2}
\eea
After inserting these expressions into equation~\eref{pnexpans} multiplied by $N$ and the results into the master equation~\eref{pi},
one may collect terms of the same order in $N^{-1/2}$. Only the terms of order $N^{-1}$ survive, leading to the partial differential equation~\eref{piw} for the joint probability density $\Pi(u,v,w)$.

\end{document}